\documentclass[%
 reprint,
  superscriptaddress,
 amsmath,amssymb,amsfonts,
 aps,
pra,
 floatfix,
 longbibliography
]{revtex4-2}
\usepackage[utf8]{inputenc}
\usepackage[pdftex]{graphicx} \graphicspath{{}}
\usepackage{float} \usepackage{color}
\usepackage[pdftex,colorlinks=true]{hyperref}
\hypersetup{
  colorlinks=true,
  linkcolor=blue,
  citecolor = blue,
  urlcolor=blue,
}

\usepackage{mathtools}

\DeclarePairedDelimiterX\braket[2]{\langle}{\rangle}{#1 \delimsize\vert #2}
\DeclarePairedDelimiterX\expval[3]{\langle}{\rangle}{#1 \delimsize\vert #2  \delimsize\vert #3}
\DeclarePairedDelimiterX\proj[2]{\delimsize\vert#1\rangle}{\langle#2\delimsize\vert}{ }
\usepackage{siunitx}

\begin{document}


\title{Localized and extended phases in square moir\'e patterns}


\author{C. Madron\~ero}
\affiliation{Instituto de F\'isica, Universidad Nacional Aut\'onoma de M\'exico, Apartado Postal 20-364, M\'exico City 01000, M\'exico}
\author{G. A. Dom\'inguez-Castro}
\affiliation{Instituto de F\'isica, Universidad Nacional Aut\'onoma de M\'exico, Apartado Postal 20-364, M\'exico City 01000, M\'exico}
\affiliation{Institut f\"ur Theoretische Physik, Leibniz Universit\"at Hannover, Appelstr. 2, D-30167 Hannover, Germany}
\author{R. Paredes}
\affiliation{Instituto de F\'isica, Universidad Nacional Aut\'onoma de M\'exico, Apartado Postal 20-364, M\'exico City 01000, M\'exico}

\date{\today}

\begin{abstract}
Random defects do not constitute the unique source of electron localization in two dimensions. Lattice quasidisorder generated from two inplane superimposed rotated, main and secondary, square lattices, namely monolayers where moir\'e patterns are formed, leads to a sharp localized to delocalized single-particle transition. This is demostrated here for both, discrete and continuum models of moir\'e patterns that arise as the twisting angle $\theta$ between main and secondary lattices is varied in the interval $[0, \pi/4]$. Localized to delocalized transition is recognized as the moir\'e patterns depart from being perfect square crystals to non-crystalline structures. Extended single-particle states were found for rotation angles associated with Pythagorean triples that produce perfectly periodic structures. Conversely, angles not arising from such Pythagorean triples lead to non-commensurate or quasidisordered structures, thus originating localized states. These conclusions are drawn from a stationary analysis where the standard IPR parameter measuring localization allowed us to detect the transition. While both, ground state and excited states were analyzed for the discrete model, where the secondary lattice was considered as a perturbation of the main one, the sharp transition was tracked back for the fundamental state in the continuous scenario where the secondary lattice is not a perturbation any more.

\end{abstract}


\maketitle

\section{Introduction}
Since the achievement of its experimental assembly, rotated graphene bilayers have attracted the attention not just because of the diversity of quantum phases accessible in them, but also by virtue of its fabrication simplicity \cite{Kim,Yankowitz,Feng,Wang,Guo}, as well as for the control achievable with respect to other multicomponent materials. The quantum phases that emerge as a result of the angle tuning include the non-conventional superconducting, metal, and Mott insulating states as a function of the carrier density \cite{Cao}, correlated ferromagnetic phases \cite{Gonzalez,Cao2,Yndurain}, and also the arrive of anomalous optical properties \cite{Chan,Chen}. Localization of charge carriers is other phenomenon that can be explored in stacked layers of graphene \cite{Trambly, Zion, Iwasaki}. In fact, it is thought that a central  mechanism affecting the electron transport in moir\'e heterostructures is the appearance of flat bands. Among other remarkable properties of electromagnetic character, is the negative magnetoresistance appearing on twisted double bilayer graphene super lattices, as a result of weak localization \cite{Kashiwagi}.

Single particle localization induced by disorder, was first explained in the seminal work of Anderson within the single electron theory \cite{Anderson}. Because of its intrinsic nature, namely, resulting from destructive wave interference, the localization phenomenon manifests not just in the degenerate quantum regime \cite{Roati, Billy}, but in any scenario in which the wave nature plays a main role \cite{Wang2}. The understanding of the opposite localized and extended states has been addresed in many different ways and schemes, both theoretically and experimentally, incorporating elements proper of the matter constituents. Among these elements it can be recognized, length and dimensionality of the lattice, inter-particle interactions \cite{Bloch, MYan, Panda, Sierant1}, long-range tunneling \cite{Santos, Modak, Modak2, Sierant2}, and of course structural disorder created either, by random potentials or quasicrystalline designs \cite{Bloch2, Dominguez-Castro1, Dominguez-Castro2}. Because of the fact that the lattice spatial dicretization must play a fundamental role, a natural question is to explore the influence of the recently created moir\'e patterns on the emergence of localized vs. extended states.
 
Ultracold bosonic atoms, largely known as quantum simulators, are schemes were the superfluid to Mott insulator transition in twisted-bilayer square lattices based on atomic Bose-Einstein condensates loaded into spin-dependent optical lattices can be tested \cite{Meng, Torma}. Several theoretic proposals difficult to realize with crystals have been planned with cold atoms to observe analogous physics to its condensed matter counterpart \cite{Tudela}.
The opportunity of preparing macroscopic clouds of weakly interacting atoms in two dimensions relies basically on two facts, one is the possibility of having a strong confinement along a spatial direction, which in turns creates a 2D Bose-Einstein condensate starting from a 3D cloud \cite{Kruger,Hadzibabic}, and the other, is the chance of varying the atom-atom interaction by tuning externally a magnetic field where the scattering length is nearly zero \cite{HungCL}, usually occurring near a Feshbach resonance. 

In this investigation we analyze the emergence of localized vs extended states of a single particle in square moir\'e lattices in two dimensions. For this purpose we consider both, a discrete lattice as well as its continuum counterpart in which a particle is confined in a two-dimensional potential composed by a principal and a secondary square lattices. Moir\'e patterns come up when the lattices lying one on top of the other are rotated by an angle $\theta$. Our analysis comprises the study of stationary localization properties.   

This paper is organized as follows. First in section II we present the model for both, continuum and discrete cases. Since the lattice model is used to analyze moir\'e patterns where the secondary lattice is considered as a perturbation, being such a perturbative term an onsite shift, a brief discusion on Wannier functions is included. In Section III we focus on the analysis of shallow moir\'e lattices and study both, ground and excited states. Then, in Section IV we concentrate in deep moir\'e patterns where the main and secondary lattices has the same depth. Finally in section V a summary of our investigation is given.

\section{Model}
\label{Model}
\begin{figure}[htbp]
\centering
\includegraphics[width=\columnwidth]{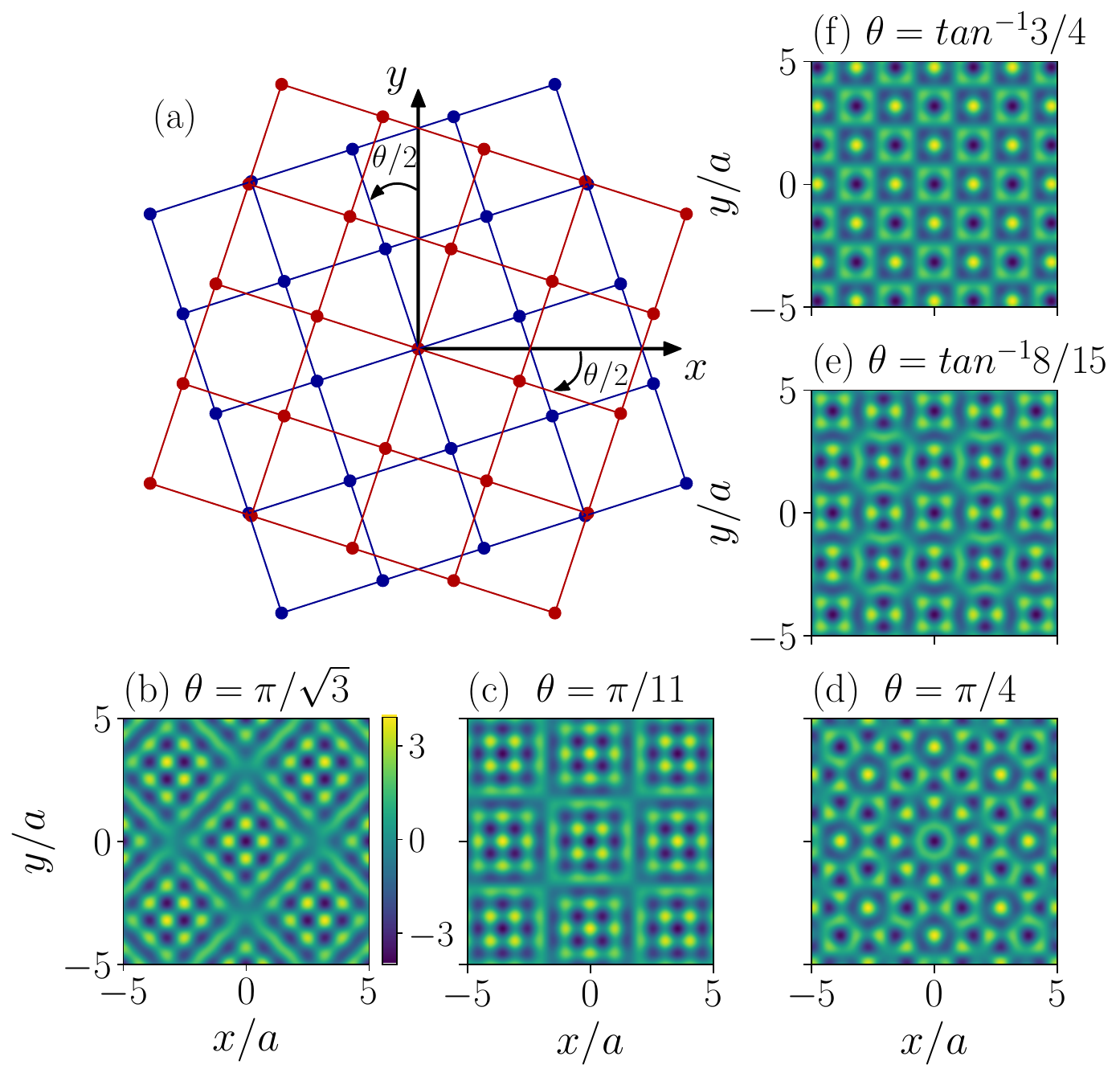}
\caption{(a) Schematic representation of two square optical lattices (blue and red) with a relative rotation angle of $\theta$ among them. Black lines correspond to the axes of an unrotated frame of reference. (b)-(f) Moir\'e patterns for several twist angles.}
\label{Figure1}
\end{figure}

Our starting point is the two-dimensional continuum single-particle Hamiltonian: 
\begin{equation}
\hat{H}_{\text{cont}} = \frac{-\hbar^{2}}{2m}\nabla_{\perp}^{2} + V_{\text{opt}}(x,y),
\label{ModelEq1}
\end{equation}
$\nabla_{\perp}^{2} = \partial_{x}^{2}+\partial_{y}^{2}$ being the Laplacian operator in 2D and $m$ the particle mass. The optical potential $V_{\text{opt}}(x,y)= V_{1}(x,y) + V_{2}(x,y)$ is given by the superposition of two square optical lattices with a relative rotation angle of $\theta$ among them. Throughout the manuscript, we call $V_{1}(x,y)$ the primary potential and $V_{2}(x,y)$ the secondary potential, each one given as follows:
\begin{equation}
\begin{split}
V_{1}(x,y) &= s_{1}E_{R}\left(\sin^{2}\frac{\pi x_{1}}{a}+\sin^{2}\frac{\pi y_{1}}{a}\right),\\
V_{2}(x,y) &= s_{2}E_{R}\left(\sin^{2}\frac{\pi x_{2}}{a}+\sin^{2}\frac{\pi y_{2}}{a}\right),
\end{split}
\label{ModelEq2}
\end{equation}
where $a$ is the lattice constant, $E_{R} = \hbar^{2}\pi^{2}/2ma^{2}$ is the recoil energy, and $s_{1}$ and $s_{2}$ are the depths of the optical potentials. In terms of an unrotated frame of reference (see Fig. \ref{Figure1}a), the coordinates $(x_{1}, y_{1})$ and $(x_{2}, y_{2})$ are given by:
\begin{equation}
\begin{split}
x_{1} &= x\cos\theta/2-y\sin\theta/2,\\
y_{1} &= x\sin\theta/2 + y\cos\theta/2,\\
x_{2} &= x\cos\theta/2+y\sin\theta/2,\\
y_{2} &= -x\sin\theta/2 + y\cos\theta/2.
\end{split}
\label{ModelEq3}
\end{equation}
Depending on the value of $\theta$, the resulting optical potential $V_{\text{opt}}(x,y)$ gives rise to periodic (commensurable) or aperiodic (incommensurable) structures. Crucially, these structures, called moir\'e lattices, always feature the rotational symmetry of the underlying sublattices except for $\theta=\pi/4$, where a quasicrystal with octagonal rotation symmetry is obtained. Commensurable moir\'e patterns result from angles that come from a Pythagorean triple, that is, $\cos \theta = m/n$, $\sin\theta = b/c$ and $m^2+n^2 = l^2$ with $m, n$ and $l$ integers. For all other rotation angles, $V_{\text{opt}}(x,y)$ leads to an incommensurable but not necessarily a disordered random lattice. In Figs. \ref{Figure1}(b)-(f), we illustrate the resulting moir\'e patterns for several rotation angles.

Formation of moir\'e patterns as a function of rotation angle gives rise to a particular spatial distribution of the local energy minima. Thus, the localized Wanneir functions, fundamental in the tight-binding approximation for nearest- or next nearest- neighbors, must be adapted for moir\'e lattices. For instance, such a picture modifies the inherent physics of band structure. Also, the transport across the lattice must be modified by the presence of the structured patterns.

\subsection{Review of basic theory for Wannier functions}
\label{Sec-Review}
\begin{figure}[htbp]
\centering
\includegraphics[width=\columnwidth]{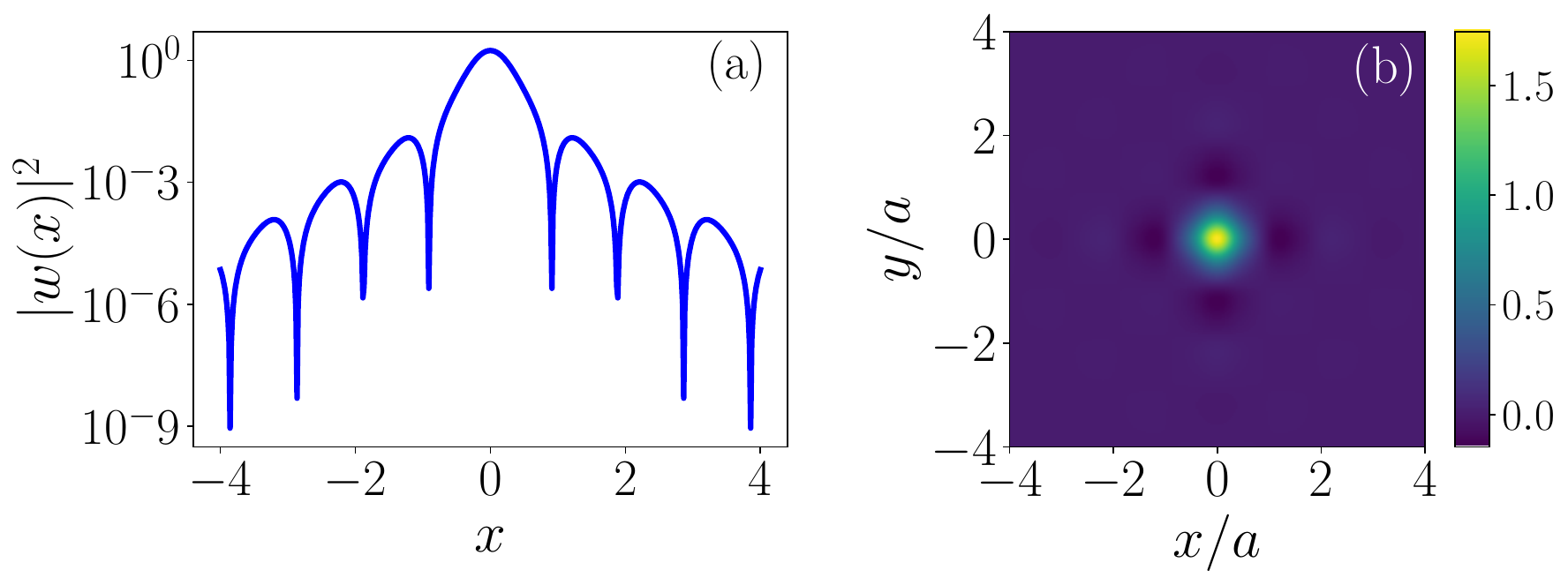}
\caption{(a) Exponential decay of a one-dimensional Wannier function, notice the log scale in the y-axis. (b) Two-dimensional Wannier function $w(x,y) = w(x)w(y)$  associated with a square lattice. }
\label{Figure0}
\end{figure}

With the purpose of establishing the discrete version of Hamiltonian (\ref{ModelEq1}), in this section we briefly review the basic concepts of a single particle subject to a periodic potential. In particular, we focus on the generation of localized Wannier functions (WFs) and their connection in the construction of a tight-binding description of a periodic Hamiltonian. A thorough discussion can be found in any standard solid state reference \cite{Vanderbilt1, Vanderbilt2, Vanderbilt3}. For simplicity, we consider a one-dimensional potential, and at the end of this section, we discuss the generalization to higher dimensions. The quantum problem of a particle of mass $m$ in a periodic potential $V(x+a) = V(x)$ with periodicity $a$ is described by the following Schr\"odinger equation: 
\begin{equation}
\left[
-\frac{\hbar^{2}}{2m}\frac{d^{2}}{dx^{2}} + V(x)
\right]\psi_{p}^{(n)}(x) = \varepsilon_{p}^{(n)}\psi_{p}^{(n)}(x),
\label{Eq1} 
\end{equation}
where $n$ labels the band number and $p\in[-\pi/a, \pi/a]$ is the quasi-momentum which is restricted to the first Brillouin zone. According to Bloch's theorem, the solutions of the above equation can be written as the product of a plane wave times a periodic function with the same periodicity of the potential, $\psi_{p}^{(n)}(x) = e^{ipx}u_{p}^{(n)}(x)$. Introducing this ansatz into Eq. (\ref{Eq1}) yields to the Schr\"odinger equation for $u_{p}^{(n)}(x)$:
\begin{equation}
\begin{split}
\left[
\frac{\hbar^{2}}{2m}\left(-i\frac{d}{dx}+p\right)^{2} + V(x) 
\right]&u_{p}^{(n)}(x) =\\ &\varepsilon_{p}^{(n)}u_{p}^{(n)}(x).
\label{Eq2} 
\end{split}
\end{equation}
The above equation can be seen as a set of eigenvalue problems, one for each $p$, with an infinite number of solutions. The collection of all eigenvalues gives rise to the corresponding band structure. Because Bloch functions extend over the entire lattice, they are not helpful for constructing a lattice Hamiltonian. Nevertheless, a convenient alternative is to use the so-called WFs. In terms of the Bloch functions, the Wannier functions are given as follows: 
\begin{equation}
w_{n}(x-x_{i}) = \frac{1}{\sqrt{L}}\sum_{p}e^{-ipx_{i}}\psi_{p}^{(n)}(x),
\label{Eq3}
\end{equation}
where $x_{i}=ia$ is the position of the $i-$th lattice site and $L$ is the number of lattice sites. A typical feature of WFs is their relatively strong concentration around each lattice minimum. Thus, they provide an attractive option for building a lattice version of a periodic Hamiltonian. Additionally, as it is easy to show, the collection of Wannier functions form an orthonormal set 
\begin{equation}
\int dx \ w_{n}^{*}(x-x_{i})w_{m}(x-x_{j}) = \delta_{n,m}\delta_{i,j}
\label{}
\end{equation}
An important subtlety of Eq. (\ref{Eq3}) is the presence of a gauge freedom that exists in the definition of Bloch functions. That is, one can replace $\tilde{\psi}_{p}^{(n)}\rightarrow e^{i\varphi_{p}}\psi_{p}^{(n)}$ without modifying the band structure. Nevertheless, such a gauge transformation will change the spatial behavior of the associated Wannier function. In one-dimension, one can always choose the phase of the Bloch waves in such a way that the corresponding WFs are real, have a well defined parity, and decay exponentially away from the central site (see Fig. \ref{Figure0}(a)). In two and three dimensions this is generally not possible \cite{Sakuma}. However, for separable potentials, the corresponding Wannier function is simply the product of the one-dimensional Wannier functions associated with each direction (see Fig. \ref{Figure0}(b)). It is important to mention that even for non-separable potentials, WFs can be generated using more elaborate procedures. Recently, Wannier functions have been generated in a quasicrystal structure, where the Bloch theorem is no longer valid \cite{Gottlob}. In section \ref{SecIII} Wannier functions will be a key element to analyze the effect of lattice localization. 

\section{Shallow moir\'e lattices}
\label{SecIII}

In this section, we study the case in which the secondary potential is weak in comparison with the primary potential, i.e. $s_{2}\ll s_{1}$. For the sake of simplicity, we consider the secondary lattice to be the only one that is rotated by an angle $\theta$. Since $s_{2}\ll s_{1}$, the minima of the main lattice are not considerably affected by the presence of the rotated lattice. Hence,  we can safely use the associated Wannier functions $w(\mathbf{r})$ of the primary lattice to build a discrete version of the Hamiltonian in Eq. (\ref{ModelEq1}). In this scenario, the lowest-band lattice Hamiltonian reads
\begin{equation}
\hat{H} = -\sum_{\mathbf{i},\mathbf{j}}J_{\mathbf{i}\mathbf{j}}(\hat{b}_{\mathbf{i}}^{\dagger}\hat{b}_{\mathbf{j}}+ h.c) + \sum_{\mathbf{i}}\epsilon_{\mathbf{i}}\hat{n}_{\mathbf{i}},
\label{Hlattice}
\end{equation}
where $\mathbf{i}=(i_{x}, i_{y})$ stands for a two-dimensional space coordinate in which $i_{x}, i_{y}$ run along the positions in a given two-dimensional lattice of size $N_{sites} = L\times L$ and $J_{\mathbf{i}\mathbf{j}}$ is the usual hopping amplitude
\begin{equation}
J_{\mathbf{i}\mathbf{j}} = \int d^{2}r \  w^{*}(\mathbf{r}-\mathbf{r}_{i})\left(-\frac{\hbar^{2}}{2m}\nabla^{2}_{\perp} + V_{1}(\mathbf{r})\right)w(\mathbf{r}-\mathbf{r}_{j})
\end{equation}
The rotated potential $V_{2}(\mathbf{r})$ can be considered  as a site-dependent energy term $\epsilon_{\mathbf{i}\mathbf{j}}$:   
\begin{equation}
\epsilon_{\mathbf{i}\mathbf{j}}= s_{2}E_{R}\int d^{2}r \ w^{*}(\mathbf{r}-\mathbf{r}_{i}) V_{\text{2}}(\mathbf{r}) w(\mathbf{r}-\mathbf{r}_{j}).
\label{epsilon}
\end{equation}
For deep enough lattices, one can rely on the so-called tight-binding approximation. In this limit, due to the exponential decay of the Wannier functions, the hopping amplitudes $J_{\mathbf{i}\mathbf{j}}$ strongly decay with the distance. Therefore, we can ignore the tunneling terms beyond the nearest neighbors. Furthermore, due to the translational symmetry of the primary lattice, the hopping parameter becomes a constant $J$ independent of the lattice site. Analogously, the leading contribution of $\epsilon_{\mathbf{i}\mathbf{j}}$ is the $\mathbf{i}=\mathbf{j}$ term, which corresponds to an on-site energy shift $\epsilon_{\mathbf{i}\mathbf{j}} = \delta_{\mathbf{i},\mathbf{j}}\ \epsilon_{\mathbf{i}}$. 
After some straightforward algebra, one can find the following expression for the on-site term:
\begin{equation}
\begin{split}
\epsilon_{\mathbf{i}} &=
\Delta[\cos2\pi(i_{x}\cos\theta+i_{y}\sin\theta)+\\
&\cos2\pi(-i_{x}\sin\theta+i_{y}\cos\theta)],
\end{split}
\label{epsilon_i}
\end{equation}
where $\Delta$ is the amplitude of the on-site potential and is given as follows:
\begin{equation}
\begin{split}
\Delta &= -\frac{s_{2}E_{R}}{2}\int dxdy \ 
|w(x,y)|^{2} \times \\
& \cos\left(\frac{2\pi x}{a}\sin\theta\right)\cos\left(\frac{2\pi y}{a}\cos\theta\right).
\end{split}
\label{Delta}
\end{equation}
\begin{figure}[htbp]
\centering
\includegraphics[width=\columnwidth]{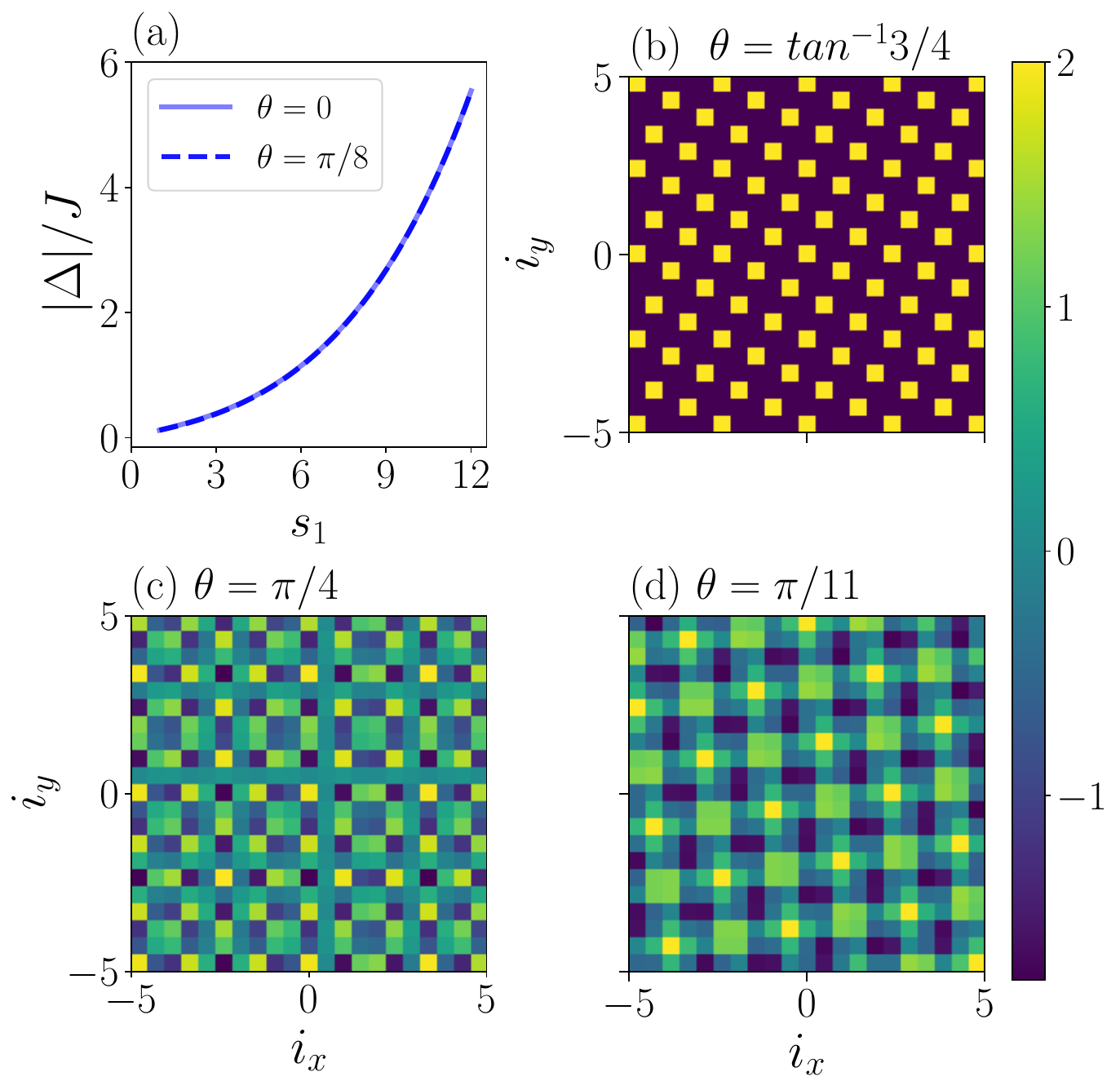}
\caption{(a) Absolute value of the on-site potential vs the primary potential depth $s_{1}$ for fixed $s_{2}=0.2$. Panels (b)-(d) illustrate the on-site potential in Eq. (\ref{epsilon_i}) as a function of the lattice coordinates for different twist angles. In all panels we fix $|\Delta|/J=1$.}
\label{Figure2}
\end{figure}

In Fig. \ref{Figure2}(a), we plot the parameter $\Delta/J$ as a function of the primary potential depth $s_{1}$ for two different twist angles. We consider a fixed value of $s_{2}=0.2$ of the depth of the secondary potential. Notice that $\Delta/J$ can be tuned to a wide range of values without changing the depth of the secondary lattice. Furthermore, the amplitude of the on-site potential depends very weakly on the angle of rotation. It is important to mention that the considered depths of the optical potentials are easily achievable in current experiments. Figs. \ref{Figure2}(b)-\ref{Figure2}(c) illustrate the on-site potential $\epsilon_{\mathbf{i}}$ as a function of the lattice coordinates for different twist angles.

\begin{figure}[htbp]
\centering
\includegraphics[width=\columnwidth]{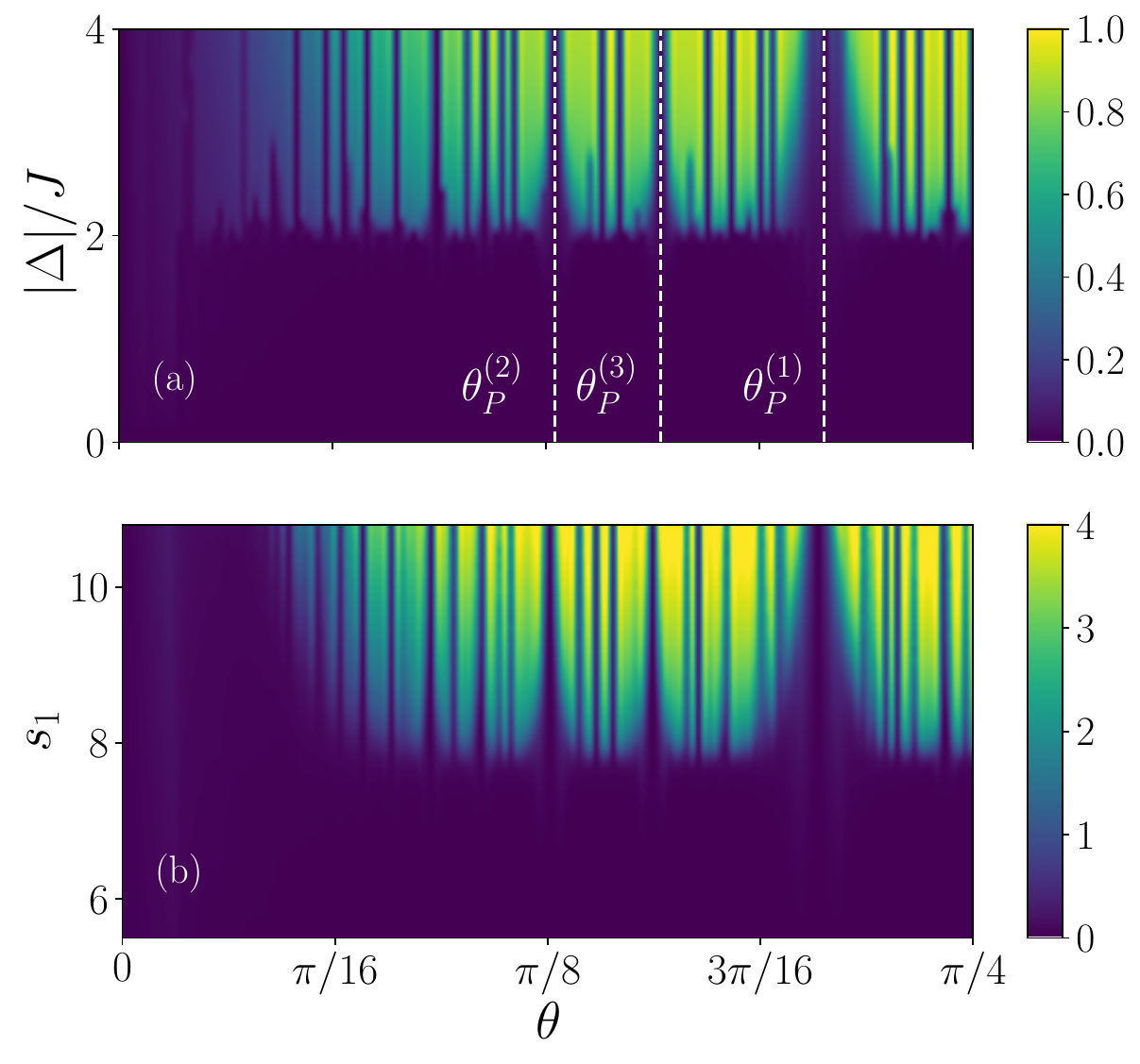}
\caption{(a) IPR associated with the ground state of the lattice Hamiltonian in Eq. (\ref{Hlattice}) as a function of the rotation angle $\theta$ and the amplitude $\Delta/J$. 
(b) Inverse participation ratio associated with the ground state of the continuum Hamiltonian in Eq. (\ref{ModelEq1}) as a function of the twist angle and the depth of the primary lattice. Both panels consider $s_{2}=0.2$.}
\label{Figure3}
\end{figure}

Having established the Hamiltonian that takes into account the secondary lattice as a perturbation term, we investigate the ground state localization properties of it. A customary parameter that is used as a measure of localization is the inverse participation ratio (IPR), given a normalized wave function $\psi$ its IPR is defined as $\text{IPR}=\sum_{i=1}^{N_{sites}}|\psi(i)|^{4}$. For extended states, the inverse participation ratio vanishes in the thermodynamic limit as $\text{IPR} \propto N_{sites}^{-1}$, whereas for localized profiles is always finite. In Fig. \ref{Figure3}(a), we plot the IPR associated with the ground state as a function of the twist angle $\theta$ and the potential strength $\Delta/J$. As shown in Fig. \ref{Figure3}(a), there is a sharp localized-delocalized transition (LDT) at $\Delta_{c}/J\simeq 2$. Below this threshold, the ground states are extended regardless of the angle of rotation. In contrast, for $\Delta/J>\Delta_{c}/J$, the spatial behavior of the the ground state becomes angle dependent. In particular, for angles given by a Pythagorean triple, identified here and henceforth as $\theta_P$, the fundamental mode is extended. In the next section we shall go back to this point.

To show the accuracy of the lattice Hamiltonian (\ref{Hlattice}), in Fig. \ref{Figure3}(b), we illustrate the IPR associated with the ground state of the continuum Hamiltonian in Eq. (\ref{ModelEq1}) as a function of the twist angle $\theta$ and the depth of the primary potential $s_{1}$. The depth of the secondary potential is considered the same as for the calculations in the lattice Hamiltonian, that is we consider $s_{2}=0.2$. Details about the numerical calculations in the continuum model can be found in Appendix \ref{Apendice-CC}. As one can notice, the results of the discrete and continuum models are in reasonable qualitative agreement, extended vs localized states appear for the same rotation angles, and definite critical values $\Delta_c/J$ and $s_1$ separates the opposite phases above $\theta \gtrsim \pi/16$. The quantitative difference between both calculations is a consequence of the different values that the IPR can take in a lattice and in continuous space, see Appendix \ref{Apendice-CC}.

\begin{figure}[htbp]
\centering
\includegraphics[width=\columnwidth]{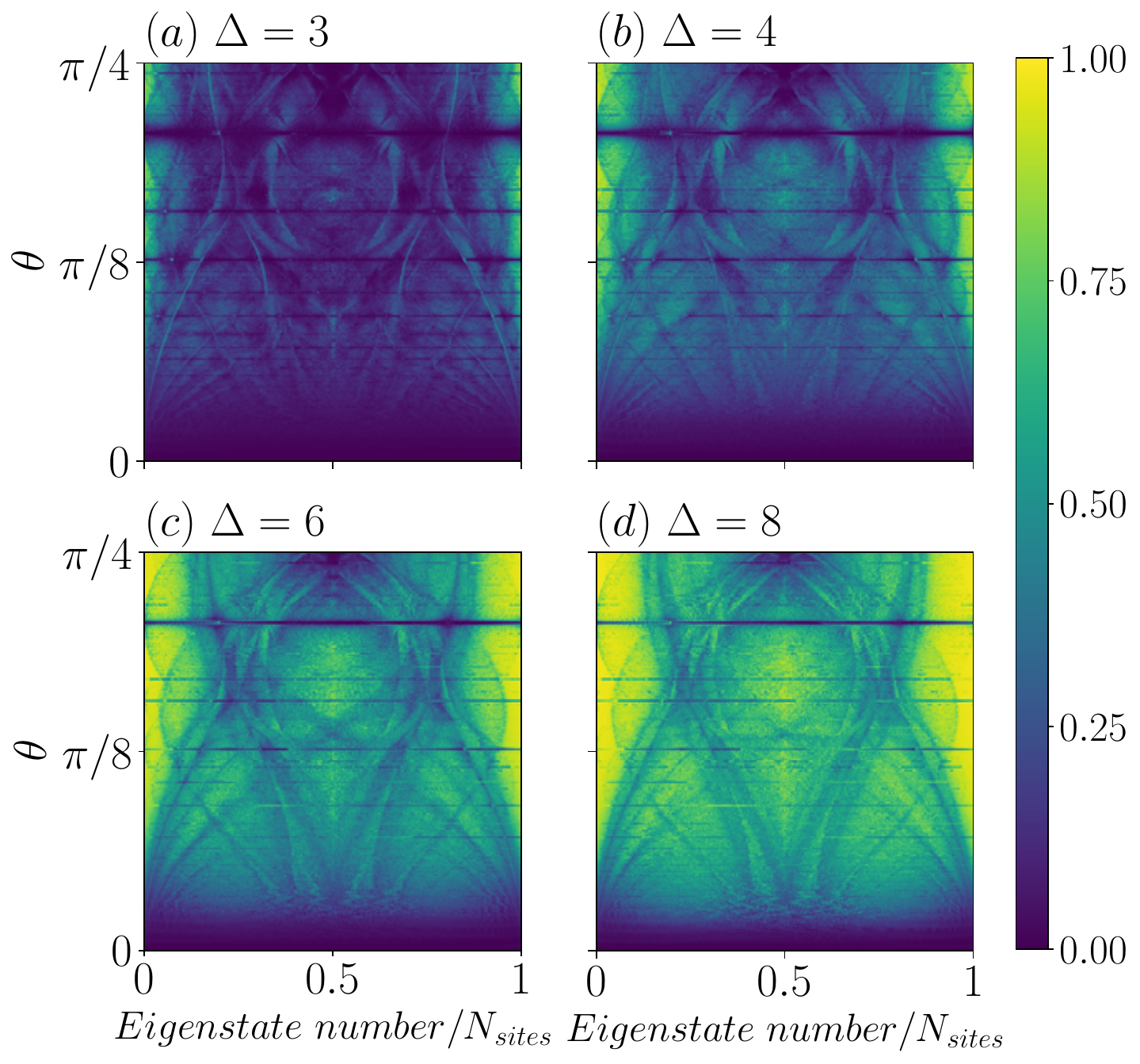}
\caption{IPR as a function of the rotation angle $\theta$ for the spectrum associated with Hamiltonian (\ref{Hlattice}).}
\label{Figure4}
\end{figure}

The building of the lattice Hamiltonian allows not only to study the localization properties of the ground state but also of the entire spectrum. In Fig. \ref{Figure4}, we plot in a color scheme the inverse participation ratio as a function of the rotation angle $\theta$ and position in the spectrum. As one can notice, the eigenstates display a rich localization structure. In particular, localized eigenstates start to appear at the edges of the spectrum leaving half of the spectrum with extended states.

\section{Deep moir\'e lattices}
In this section, our focus is on the case where the potential amplitude of the secondary lattice is comparable or larger than that of the main lattice. In this particular scenario, the secondary potential ceases to be a mere perturbation, leading to a significant impact on the minima of the principal potential. Consequently, the Wannier functions associated with the principal potential are no longer suitable for constructing a lattice Hamiltonian. 

\begin{figure}[htbp]
\centering
\includegraphics[width=\columnwidth]{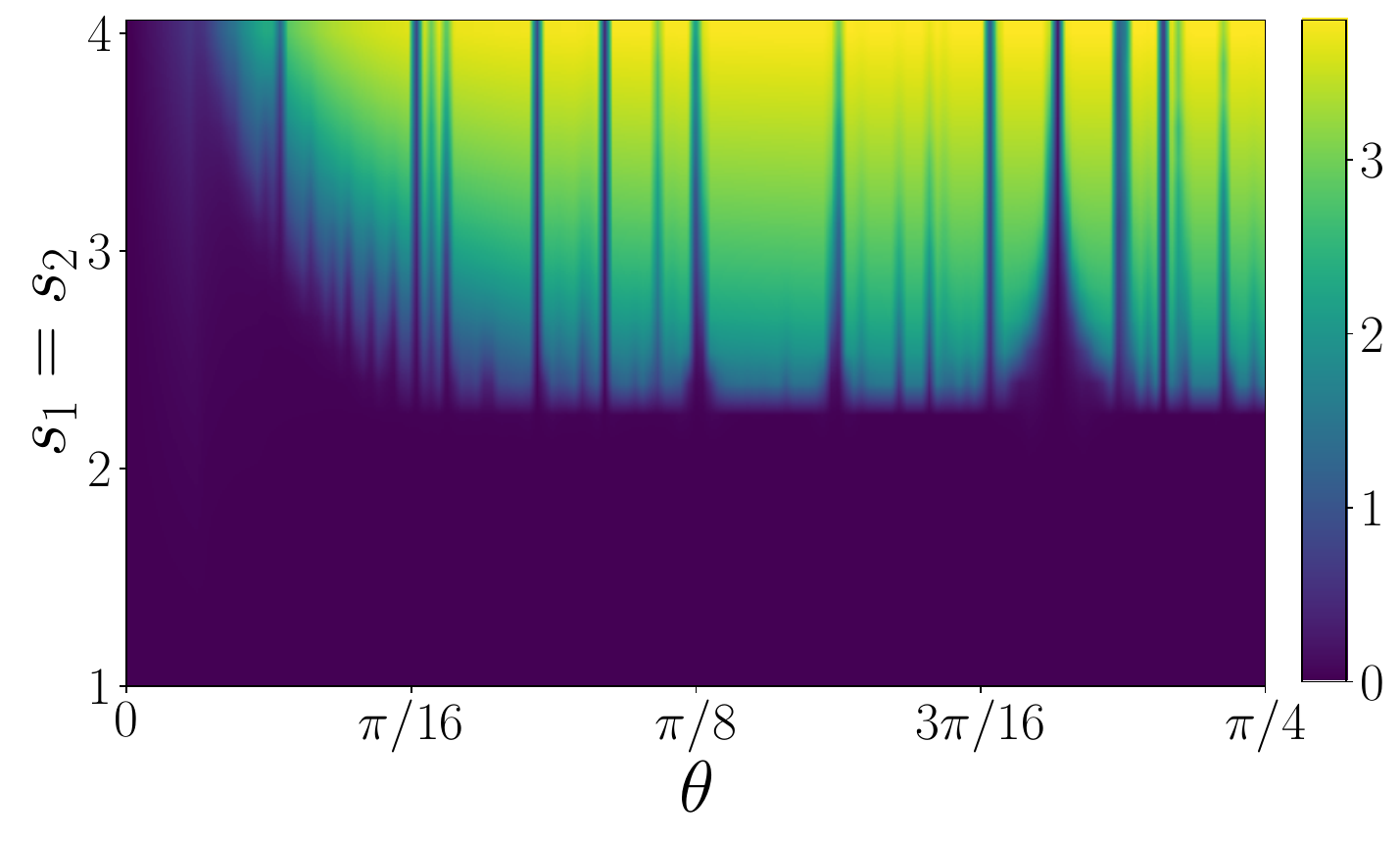}
\caption{IPR as a function of the rotation angle $\theta$ for the ground state of Hamiltonian (\ref{ModelEq1}). The values of the potential depths in main and secondary lattices are equal (see Eq. (\ref{ModelEq2})), $s_1=s_2$.}
\label{IPRs1-s2}
\end{figure}

For the analysis we solve the continuous Schr\"odinger equation for Hamiltonian (\ref{ModelEq1}) considering lattices having $N_{sites} = 50\times 50$. As in previous section, we investigate the IPR as a function of the rotation angle for the ground state. In figure \ref{IPRs1-s2} we plot in a density color scheme the IPR as a function of both, the rotation angle and the lattice depth $s_1=s_2$. As can be apreciated from this figure, the behavior is similar, qualitatively, to that found for shallow moir\'e lattices analyzed in the previous section. Starting from a certain angle, a sharp LDT for the ground state again occurs at a given potential amplitude. Instead of finding the transition at $s_1 \approx 8$ as in the shallow moir\'e lattice case, one observes that the LDT appears at $s_1 \approx 2.0$. The behavior of the IPR for small angles near the LDT, say $\theta \lesssim \pi/16$, signals the presence of an extended ground state disregarding the twisting angle (see dark blue region). The dark blue region (for small rotation angles) suggesting extended states nearly above $s_1 \approx 8$ for shallow moir\'e lattices, and $s_1 \approx 2.0$ for deep moir\'e lattices, is, as we argue below, a result of the impossibility of exploring moir\'e lattices arising from small rotation angles. In other words, limitations imposed by numerical calculations for small twisting  angles prevent us to have reliable predictions. Certainly, as the potential depth grows, it is expected that the ground state be a localized one because tunneling across sites must suffer a reduction.

\begin{figure}[htbp]
\centering
\includegraphics[width=\columnwidth]{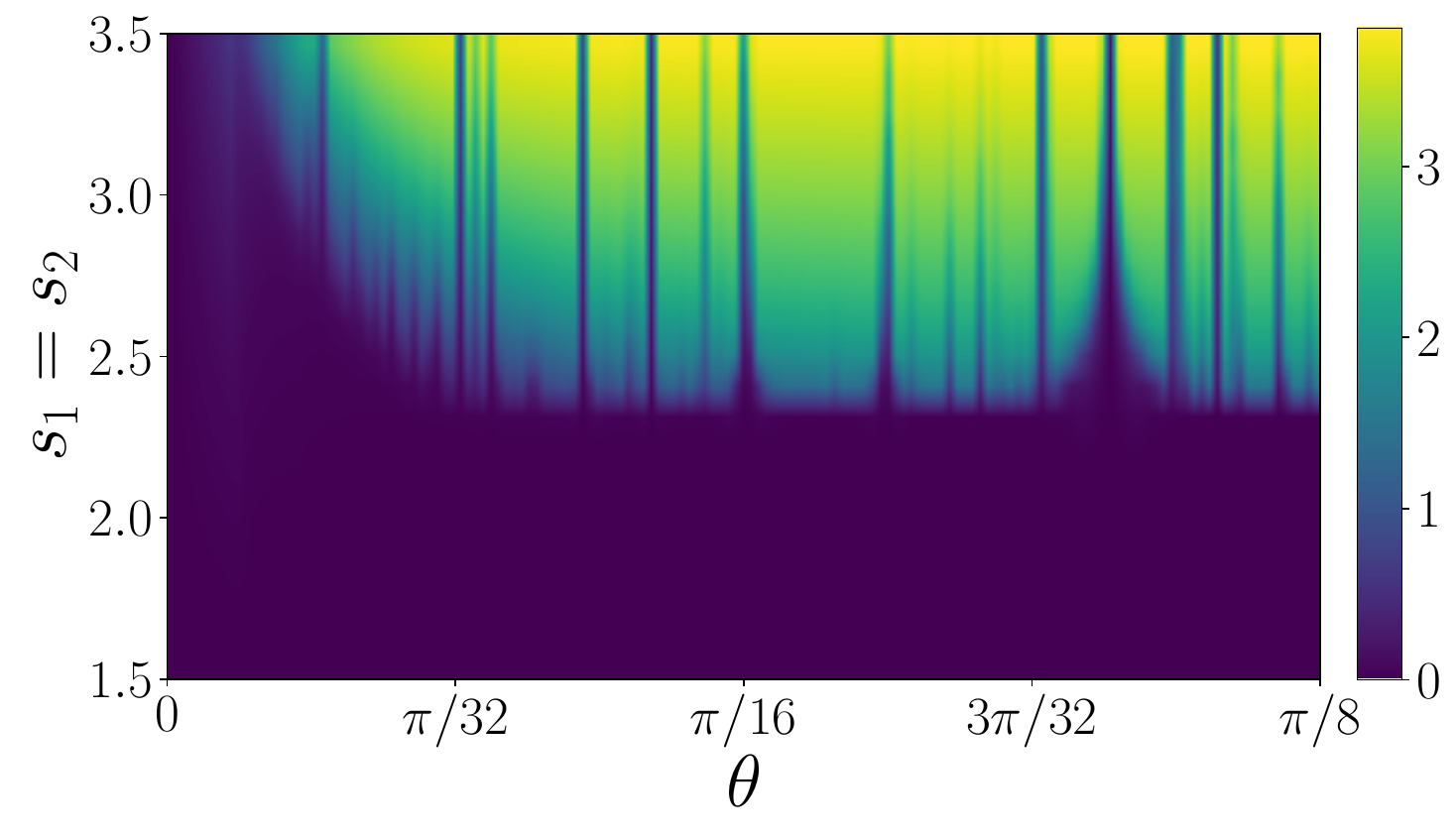}
\caption{IPR as a function of the rotation angle $0< \theta < \pi/8$ for the ground state of Hamiltonian (\ref{ModelEq1}). The values of the potential depths in main and secondary lattices are equal (see Eq. (\ref{ModelEq2})), $s_1=s_2$. The lattice has $N_{sites}=110 \times 110$ sites.}
\label{Fignew}
\end{figure}

\begin{figure}[htbp]
\centering
\includegraphics[width=\columnwidth]{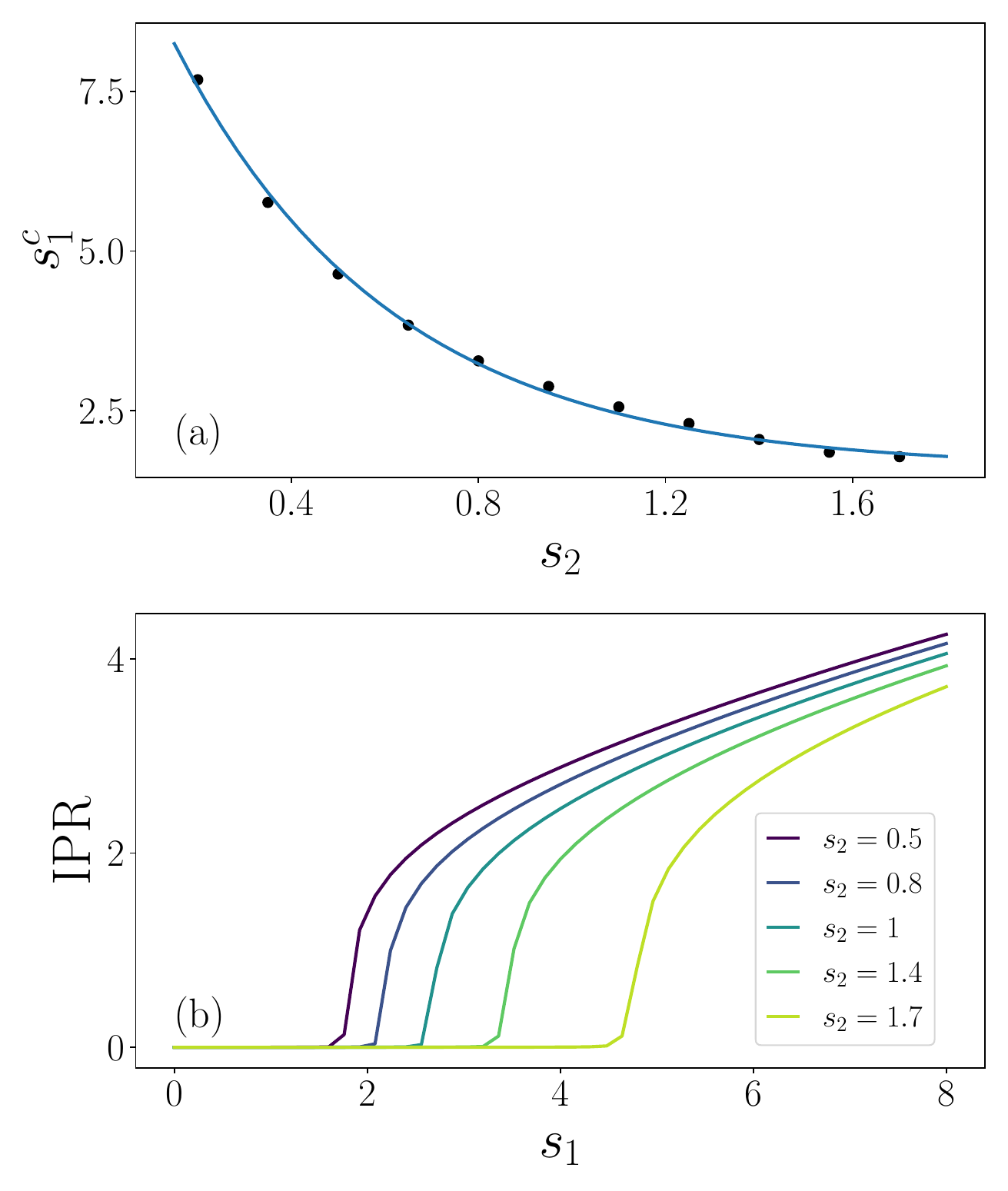}
\caption{In (a) black dots are the critical value of the amplitude $s_1^c$ that signals the LDT vs. the amplitude of the secondary lattice amplitude appears. Blue curve in this inset fits the black dots. (b) IPR as a function of potential depth $s_1$ for several laticce sizes is plotted. The lattice size used is $N_{sites}= 50 \times 50$.}
\label{Fignew2}
\end{figure}

As described previously, rotation angles associated with Pythagorean triples $\theta_P$ produce crystalline moir\'e patterns, being the lattice constant of those periodic arrays given by $a_{MC}= a/\sin{\theta}$ \cite{Feuerbacher,CMadronero}, where $a$ is the lattice constant of the main square lattice. Since the lattice constant grows as the rotation angle becomes smaller, a larger number of sites is needed to allow an appropriate exploration of the lattice space, and thus determine if the fundamental state is localized or extended. We analyzed lattices having a larger number of sites, as well as considering a smallest grid for the rotation angle between main and secondary lattices. In figure \ref{Fignew} we plot the IPR for $\theta < \pi/8$ for 2D lattices having $N_{sites}=110 \times 110$ sites. As can be seen from this figure, several new sharp regions appear, indicating extended states not advised previously. These extended states correspond to other Pythagorean triads as described above. It appears in figure \ref{Fignew} that the LDT occurs at $s_1=s_2 \approx 1.75$, but such an appearance must be attributed to the color scheme, as can be appreciated from the color bar. In figure \ref{Fignew2} we illustrate the LDT for as a function of the potential depth $s_2$. The points correspond to the value of $s_1$ at which the ground state becomes localized, such a critical value was chosen as the point at which the IPR is no longer zero. The blue line corresponds to an exponential fit $s_1^c=Ae^{-B s_2 }+C$ \cite{Fit}. At the inset of this figure we include several IPR curves associated with given values of $s_2$ indicated in colors.

\begin{figure}[htbp]
\centering
\includegraphics[width=\columnwidth]{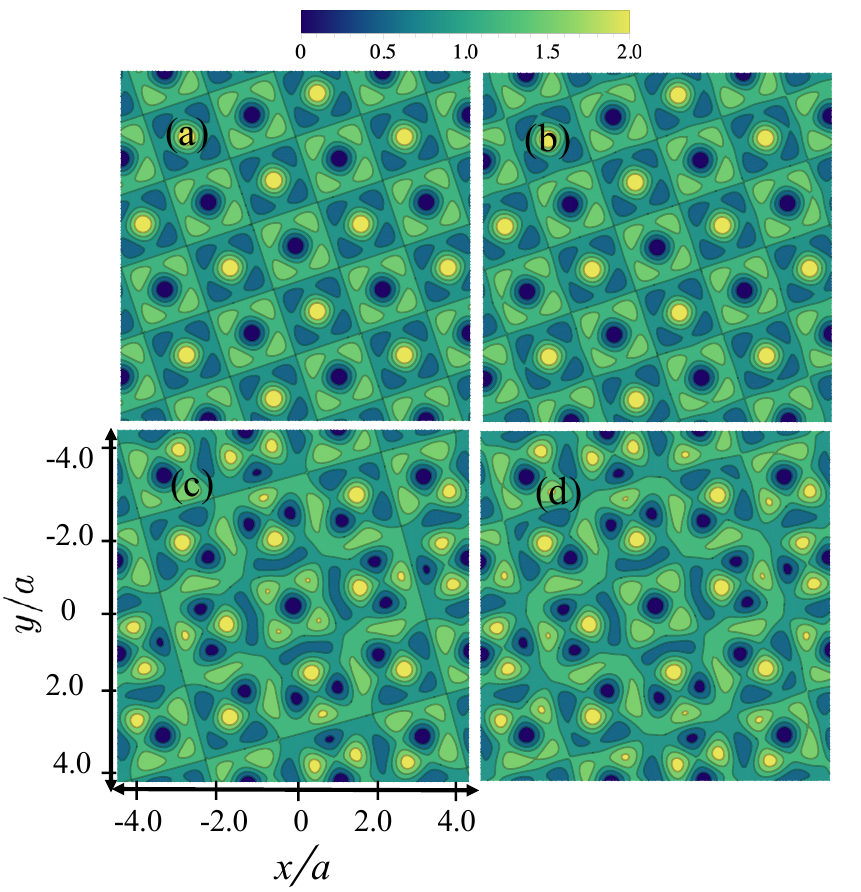}
\caption{Contour curves of potential in Eq. (\ref{ModelEq2}) for a) $\theta_P^{(1)}=\arctan 3/4$, (b) $\theta=\arctan 3/4 +0.01$, (c) $\theta_P^{(8)}=\arctan 33/56$, and (d) $\theta=\arctan 33/56 +0.01$.  The values of the potential depths are the same for each panel $s_1=s_2=1.5$}
\label{PotLand}
\end{figure}

The reason behind the appearance of extended states for Pythagorean triples, namely crystalline moir\'e structures, is a direct consequence of its periodicity. As stated by the Bloch theorem, all the single-particle eigenstates are extended, particularly the ground state. Thus one reaches the conclusion that all of the moir\'e lattices coming from rotation angles associated with Phytagorean triples lead to extended states. It is worth to mention here that for $\theta \in [0, \pi/4)$ one can find an infinite number of Phytagorean triples \cite{JFallas}. The Pythagorean angles appearing in figures \ref{Figure3} and \ref{IPRs1-s2} are those captured by our numerical grid. Conversely, deviation of commensurate structures, leads to obtain localized states above a certain treshold value of $s_1$. This result is reminiscent of the findings exhibited by the Aubry-Andr\'e model for quasicrystalline lattices, which establishes that above a critical disorder strenght all the states are localized, and in particular the ground state \cite{Dominguez-Castro3}. This behavior can be understood from the potential landscape where the single-particle moves. As shown in figure \ref{PotLand}, non-crystalline structures are such that consequtive potential minima are quite appart with respect to those associated with perfectly commensurate lattices (see figures \ref{PotLand}(c) and \ref{PotLand}(d)). In summary, a moir\'e lattice that departs from a perfect crystal, and becomes a non periodic structure, shows in good agreement with the seminal result of Anderson, localized states starting from a certain threshold. At this point one can mention that our quasidisordered moir\'e lattices do not show critical states as those exhibited in the 2D generalized Aubry-Andr\'e model \cite{Duncan}.  

Not less important is the particular feature exhibited by certain twisting angles generating crystalline moir\'e lattices. As can be seen from figures \ref{Figure3} and \ref{IPRs1-s2}, the white line indicating $\theta_P^{(1)}=\arctan 3/4$ is the center of a wider region signaling extended states, in comparison with other angles associated to different Pythagorean triads, as $\theta_P^{(2)} =\arctan 5/12$ and $\theta_P^{(3)}=\arctan 8/15$, for which the region of extended states is narrower. The explanation is the following. The minima of moir\'e crystals, that is the spatial regions where the potential $V_{\text{opt}(x,y)}$ takes its minumum values, are separated among them by a distance determined by the Pythagorean triad that defines a given structure. Such a separation is the hypotenuse of the right triangle $m^2+n^2=l^2$, being $m$, $n$ and $l$ integer multiples of the lattice constant $a$. Some examples of these Pythagorean triples and its associated angles are $\theta_P^{(1)} \to (3,4,5)$, $\theta_P^{(2)} \to (5,12,13)$, $\theta_P^{(2)} \to (8,15,17)$, and $\theta_P^{(11)}\to (33,56,65)$, where the label in parenthesis has been asigned considering the increasing  size of the hypotenuse. Let us consider the origin and the rotation center coincide with one of the potential minima associated to a given Pythagorean angle $\theta_P$. When a small rotation around $\theta_P$ is performed, namely $\theta_P \pm \delta \theta$, the nearest potential minima apart each other by a distance $ \sim \pm l \delta \theta$. Therefore, for the same $\delta \theta$, the minima will be further away as the hypotenuse grows, which breaks the potential periodicity. In the neigborhood of the rotation center appreciable variations of the potential can be appreciated. This is illustrated in a density color scheme in Figure \ref{PotLand}. There, we show the contour curves of $V_{\text{opt}}(x,y)$ of Eq. (\ref{ModelEq2}) for a couple of angles $\theta_P$. As can be seen from panels in this figure, detectable changes are seen between (c) and (d) corresponding to $\theta_P^{(11)}= \arctan 33/56$ and $\theta= \arctan 33/56+0.01$ respectively, while imperceptible changes appear for the cases $\theta_P^{(1)}= \arctan 3/4$ and $\theta= \arctan 3/4+0.01$ illustrated in figures (a) and (b) respectively.

\section{Conclusions}

This investigation has deal with the identificaction of extended vs. localized phases of a single particle confined in two superimposed square lattices rotated one with respect to the other by an angle in the interval $(0, \pi/4]$. The lattices lying in the same plane are such that the amplitude of the secondary lattice can be either, shallow or equal to that of the main lattice. The patterns that result from the superposition, the so called moir\'e structures, can be classified in two types; crystalline and quasidisorderdered lattices. While the former result from twisting angles associated with the so called Pythagorean triples (satisfying a diphantine equation) and lead to perfect commensurability among main and secondary lattices, the later correspond to arbitrary angles and produce non-commensurate or quasidisordered structures. As described below, the structure plays a crucial role on the localized to delocalized transition.

Localized and delocalized phases were detected from stationary properties, considering two approaches: the discrete or lattice model, and the continuous one. In the case of the lattice model we constructed a Hamiltonian that considers the secondary lattice as an onsite perturbation term, obtained from the Wannier functions of the main lattice. Localized to delocalized transition was tracked from the standard IPR parameter as a function of the rotation angle and the potential depth of main and secondary laticces. Properties of both, fundamental and excited states were investigated for the lattice model. Regarding the continuous model, we concentrate in studying the ground state for both, shallow and deep moir\'e lattices. 

The information provided by the IPR parameter leads us to reach the following conclusions. Extended phases emerge for crystalline moir\'e structures, while localized ones are identified for non-commensurate structures starting from certanin potential depths. The identification of such a critical potential amplitude, for both shallow and deep moir\'e lattices, at which a sharp localized-delocalized transitiosn occurs suggest that moir\'e patterns are the generalization of the one-dimensional Aubry-Andr\'e model. Finaly, it is worth to mention that the existence of particular twisting Pythagorean angles for which robust extended states emerge, as for instance $\theta= \arctan 3/4$. As the Pythagorean triad associated with a given structure is such that the hypotenuse becomes larger, the narrower becomes the line signaling the extended phase.  

Similarly to light propagation phenomena in moir\'e arrays, one finds that matter confined in moir\'e patterns offers also new possibilities regarding the transport and localization properties. For instance, moir\'e heterostructures greatly enhance the appearance of new types of phenomena that are under current investigation in 2D materials.

\section*{Acknowledgements}
This work was partially funded by Grant No. IN108620 from DGAPA (UNAM). C.J.M.C acknowledges CONACYT scholarship. G.A.D.-C. acknowledge support of the Deutsche Forschungsgemeinschaft (DFG, German Research Foundation) under Germany’s Excellence Strategy – \ EXC-2123 QuantumFrontiers \ – \ 390837967.

\appendix

\section{Continuum Calculations}
\label{Apendice-CC}
In this appendix, we provide further details on the continuum calculations. For the ground state and determination of IPR we consider the parameters that appear in the following tables. 

\begin{table}[H]
\centering
\small
\begin{tabular}{p{4.5cm} c  c} 
Name & Symbol & Value\\
\hline\hline
Number of grid points in the $x$ direction & $N_{x}$ & 512-2048\\
Number of grid points in the $y$ direction & $N_{y}$ & 512-2048\\
Spatial extension of the numerical grid in the $x$ direction & $L_{x}$ & 50-200 $a$ \\
Spatial extension of the numerical grid in the $y$ direction  & $L_{y}$ & 50-200 $a$ \\
Step size used in real time evolution & $d\tau$ & $0.005$\\
\end{tabular}
\caption{Parameters for the numerical simulation}
\label{Table1}
\end{table}

\begin{table}[h]
\centering
\small
\begin{tabular}{p{4.5cm} c  c} 
Name & Symbol & Value\\
\hline\hline
Lattice constant & $a$ & $532$ nm\\
Potential depth  & $V_{0}$ & 0-11 $E_R$  \\
\end{tabular}
\caption{Physical parameters used in the numerical simulation}
\label{Table2}
\end{table}

A brief comment regarding our numerical calculations is in order. While the number of commensurate lattices that arise from two rotated square lattices is infinite, namely angles originating crystalline structures, numerical restrictions impede both, to employ lower angle grids, and the use of larger lattices.



\begin{thebibliography}{9}


\bibitem{Kim}  K. Kim, M.A. Yankowitz, B. Fallahazad, S. Kang, and E. Tutuc,
\href{https://pubmed.ncbi.nlm.nih.gov/26859527/} {Nano Lett. 16(3), 1989 (2016). } 

\bibitem{Yankowitz} M. Yankowitz et. al.,
\href{https://www.science.org/doi/10.1126/science.aav1910} {Science {\bf 363} (6431),1059 (2019).}


\bibitem{Feng}  X. Feng, S. Kwon, J.Y. Park, and M. Salmeron,
\href{https://pubs.acs.org/doi/10.1021/nn305722d}{ACSNano {\bf 7} (2),1718  (2013).}

\bibitem{Wang} D. Wang et. al., 
\href{https://journals.aps.org/prl/abstract/10.1103/PhysRevLett.116.126101}{Phys. Rev. Lett. {\bf 116} (12),126101 (2016)}

\bibitem{Guo} H.-W. Guo, Z. Hu, Z.-B. Liu, J.-G. Tian, 
\href{https://onlinelibrary.wiley.com/doi/abs/10.1002/adfm.202007810}{Adv. Func. Mater. {\bf 31} (4), 2007810 (2021). }

\bibitem{Cao} Y. Cao, V. Fatemi, A. Demi, S. Fang, S. Tomarken, J. Y. Luo, J. Sanchez-Yamagishi, K. Watanabe, T. Taniguchi, E. Kaxiras, and P. Jarillo-Herrero, 
\href{https://www.nature.com/articles/nature26160}{Nature {\bf 556}, 43 (2018).}

\bibitem{Gonzalez} L. A. Gonz\'alez-Arraga, J. L. Lado, F. Guinea, and P. San-Jose,  
\href{https://journals.aps.org/prl/abstract/10.1103/PhysRevLett.119.107201}{Phys. Rev. Lett. {\bf 119}, 107201 (2017).}

\bibitem{Cao2} Y. Cao, V. Fatemi, A. Demi, S. Fang, S. Tomarken, J. Y. Luo, J. Sanchez-Yamagishi, K. Watanabe, T. Taniguchi, E. Kaxiras, et al.,  
\href{https://www.nature.com/articles/nature26154}{Nature {\bf 556}, 80 (2018).}

\bibitem{Yndurain} F. Yndurain, 
\href{https://journals.aps.org/prb/abstract/10.1103/PhysRevB.99.045423}{Phys. Rev. B {\bf 99}, 045423 (2019).}

\bibitem{Chan} M.-Ch. Chan, Y.-Ch. Chen, B.-H. Shiue, T.-I Tsai, Ch.-D. Chen, and W.-S. Tseng, \href{https://opg.optica.org/oe/fulltext.cfm?uri=oe-29-24-40481&id=464991}{ Opt. Express {\bf 29}, 40481 (2021).}

\bibitem{Chen} Z. Chen, H. Yang, Y. Xiao, J. Pan, Y. Xia, and W. Zhu, Tseng, 
\href{https://opg.optica.org/josaa/abstract.cfm?uri=josaa-38-8-1232}{J. Opt. Soc. Am. A {\bf 38}, 1232 (2021). }

\bibitem{Trambly} G. Trambly de Laissardiere, D. Mayou, and L. Magaud,
\href{https://pubs.acs.org/doi/pdf/10.1021/nl902948m}{Nano Lett. {\bf 10} 804 (2010).} 

\bibitem{Zion} E. Zion, A. Haran, A. V. Butenko, L. Wolfson, Y. Kaganovskii, T. Havdala1, A. Sharoni, D. Naveh, V. Richter, M. Kaveh, E. Kogan, I. Shlimak,
\href{https://www.scirp.org/pdf/Graphene_2015072814322479.pdf} {Graphene {\bf 4} 45 (2015).}

\bibitem{Iwasaki} T. Iwasaki, S. Moriyama, N. F. Ahmad, K. Komatsu, K. Watanabe, T. Taniguchi, Y. Wakayama, A. M. Hashim, Y. Morita and S. Nakaharai,
\href{https://www.nature.com/articles/s41598-021-98266-4} {Sci.Rep. {\bf 11} 18845 (2021).}


\bibitem{Kashiwagi} M. Kashiwagi, T. Taen, K. Uchida, K. Watanabe, T. Taniguchi, and T. Osada,
\href{https://iopscience.iop.org/article/10.35848/1347-4065/ac934a/pdf}{Japanese Journal of Applied Physics {\bf 61}, 100907 (2022). } 

\bibitem{Anderson} P. W. Anderson, 
\href{https://journals.aps.org/pr/abstract/10.1103/PhysRev.109.1492}{Phys. Rev. {\bf 109}, 1492 (1958).}

\bibitem{Roati} G. Roati, C. D'Errico, L. Fallani, M. Fattori, C. Fort, M. Zaccanti, G. Modugno, M. Modugno, and M. Inguscio,
\href{https://www.nature.com/articles/nature07071}{Nature (London) {\bf 453}, 895-898 (2008). }

\bibitem{Billy} J. Billy, V. Josse, Z. Zuo, A. Bernard, B. Hambrecht, P. Lugan, D. Cl\'ement, L. Sanchez-Palencia, P. Bouyer and A. Aspect, 
\href{https://www.nature.com/articles/nature07000}{Nature (London) {\bf 453}, 891-894 (2008).}

\bibitem{Wang2} P. Wang, Y. Zheng, X. Chen, C. Huang, Y. V. Kartashov, L. Torner, V. V. Konotop and F. Ye,
\href{https://www.nature.com/articles/s41586-019-1851-6}{Nature {\bf 577}, 42-46 (2020).}

\bibitem{Bloch} J.-Y. Choi et. al. \href{https://www.science.org/doi/10.1126/science.aaf8834} {Science {\bf 352}, 1547-1552 (2016).}

\bibitem{MYan} M. Yan, H.-Y. Hui, M. Rigol, and V. W. Scarola,
\href{https://journals.aps.org/prl/abstract/10.1103/PhysRevLett.119.073002}{Phys. Rev. Lett. {\bf 119}, 073002 (2017)}.

\bibitem{Panda} R. K. Panda, A. Scardicchio, M. Schulz, S. R. Taylor, and M. {\v{Z}}nidari{\v{c}},
\href{https://iopscience.iop.org/article/10.1209/0295-5075/128/67003}{EPL (Europhysics Letters) {\bf 128}, 67003 (2019)}. 

\bibitem{Sierant1} P. Sierant and J. Zakrzewski, \href{https://journals.aps.org/prb/pdf/10.1103/PhysRevB.105.224203}{Phys. Rev. B {\bf 105}, 224203 (2022)}.

\bibitem{Santos} X. Deng, S. Ray, S. Sinha, G. V. Shlyapnikov, and L. Santos, 
\href{https://journals.aps.org/prl/abstract/10.1103/PhysRevLett.123.025301}{Phys. Rev. Lett. {\bf 123}, 025301 (2019)}.

\bibitem{Modak} R. Modak and T. Nag, \href{https://journals.aps.org/prresearch/abstract/10.1103/PhysRevResearch.2.012074}{Phys. Rev. Research {\bf 2}, 012074(R) (2020)}.

\bibitem{Modak2} R. Modak and T. Nag,
\href{https://journals.aps.org/pre/abstract/10.1103/PhysRevE.101.052108}{Phys. Rev. E {\bf 101}, 052108 (2020)}.

\bibitem{Sierant2} P. Sierant, E. G. Lazo, M. Dalmonte, A. Scardicchio, and J. Zakrzewski,
\href{https://journals.aps.org/prl/abstract/10.1103/PhysRevLett.127.126603}{Phys. Rev. Lett. {\bf 127}, 126603 (2021)}.

\bibitem{Bloch2} M. Schreiber et. al.  \href{https://www.science.org/doi/10.1126/science.aaa7432}{Science {\bf 349}, 842-845 (2015)}.

\bibitem{Dominguez-Castro1} G. A. Dom\'inguez-Castro and R. Paredes,
\href{https://journals.aps.org/pra/abstract/10.1103/PhysRevA.104.033306}{Phys. Rev. A {\bf 104}, 033306 (2021)}.

\bibitem{Dominguez-Castro2} G. A. Dom\'inguez-Castro and R. Paredes,
\href{https://journals.aps.org/prb/abstract/10.1103/PhysRevB.106.134208}{Phys. Rev. B {\bf 106}, 134208 (2022)}.

\bibitem{Meng} Z. Meng, L. Wang, W. Han, F. Liu, K. Wen, Ch. Gao, P. Wang, Ch. Chin and J. Zhang, \href{ https://doi.org/10.1038/s41586-023-05695-4}{ Nature, {\bf 615}, 231-236, (2023).} 

\bibitem{Torma} P. T\"orma, S. Peotta and B. Bernevig,
\href{https://www.nature.com/articles/s42254-022-00466-y}{Nature Reviews, {\bf 4}, 528-542  (2022).} 

\bibitem{Tudela} A. Gonz\'alez-Tudela and J. I. Cirac,
\href{https://journals.aps.org/pra/abstract/10.1103/PhysRevA.100.053604}{Phys. Rev. A {\bf 100}, 053604 (2019).} 

\bibitem{Kruger} P. Kr\"uger, Z. Hadzibabic, and J. Dalibard, 
\href{https://journals.aps.org/prl/abstract/10.1103/PhysRevLett.99.040402}
{Phys. Rev. Lett. {\bf 99}, 040402 (2007).} 

\bibitem{Hadzibabic} Z. Z. Hadzibabic, P. Kr\"uger, M. Cheneau, B. Battelier and J. Dalibard, 
\href{https://pubmed.ncbi.nlm.nih.gov/16810249/}
{Nature {\bf 441} 1118 (2006)}.

\bibitem{HungCL} C.-L. Hung, X. Zhang, N. Gemelke and C. Chin, \href{https://www.nature.com/articles/nature09722}  {Nature {\bf 470}, 236 (2011).} 

\bibitem{Vanderbilt1} N. Marzari, D. Vanderbilt, \href{https://journals.aps.org/prb/abstract/10.1103/PhysRevB.56.12847} {Phys. Rev. B  {\bf 56}, 12847 (1997).} 

\bibitem{Vanderbilt2} I. Souza, N. Marzari, D. Vanderbilt, \href{https://journals.aps.org/prb/abstract/10.1103/PhysRevB.56.12847} {Phys. Rev. B  {\bf 65}, 035109 (2001).} 

\bibitem{Vanderbilt3} N. Marzari, A.A. Mostofi, J.R. Yates, I. Souza, D. Vanderbilt, \href{https://journals.aps.org/rmp/abstract/10.1103/RevModPhys.84.1419} {Rev. Mod. Phys.  {\bf 84}, 1419 (2012).} 

\bibitem{Sakuma} R. Sakuma, 
\href{https://journals.aps.org/prb/pdf/10.1103/PhysRevB.87.235109}{Phys. Rev. B {\bf 87}, 235109 (2013). }

\bibitem{Gottlob} E. Gottlob and U. Schneider, 
\href{https://journals.aps.org/prb/pdf/10.1103/PhysRevB.107.144202}{Phys. Rev. B {\bf 107}, 144202 (2023).}

\bibitem{Feuerbacher} M. Feuerbacher,
\href{https://scripts.iucr.org/cgi-bin/paper?S2053273321007245}{Acta Cryst. {\bf A77}, 460 (2021).}


\bibitem{CMadronero} C. Madro\~nero and R. Paredes, 
\href{https://journals.aps.org/pra/abstract/10.1103/PhysRevA.107.033316}{Phys. Rev. A, {\bf 107}, 033316 (2023).} 


\bibitem{Fit} The values of the fit paremeters are: $A= 9.186$, $B=2.154$ and $C=1.594$.

\bibitem{JFallas} J. J. Fallas,
\href{http://funes.uniandes.edu.co/8080/2/Fallas2009Ternas.pdf}{Revista digital Matemática, Educación e Internet, {\bf 9}, 2, (2009)}.

\bibitem{Dominguez-Castro3} G.A.Dom\'{\i}nguez-Castro and R. Paredes, \href{https://iopscience.iop.org/article/10.1088/1361-6404/ab1670}{European Journal of Physics, {\bf 40}, 045403, (2019).}


\bibitem{Duncan} C.W. Duncan, \href{https://journals.aps.org/prb/abstract/10.1103/PhysRevB.109.014210}{Phys. Rev. B {\bf 109}, 014210 (2024)}


\end{thebibliography}
\end{document}